\documentclass[%
 reprint,
 superscriptaddress,
 amsmath,amssymb,
 aps,
]{revtex4-2}

\usepackage{url}
\usepackage{graphicx}
\usepackage{bm}
\usepackage{hyperref}
\usepackage[autostyle]{csquotes}
\usepackage{booktabs}
\usepackage{physics}
\usepackage{mathtools}
\usepackage{comment}

\begin{document}






\title{Spatially Dense, Continuous-Variable Quantum Computing with Solid State Spin Nonlinearities}

\author{Hamza Raniwala}
\affiliation{Department of Electrical Engineering and Computer Science, Massachusetts Institute of Technology, Cambridge, MA 02139, USA}

\author{Ethan G Arnault}
\affiliation{Institute for Soldier Nanotechnologies, Massachusetts Institute of Technology, Cambridge, MA 02139, USA}

\author{Dirk R. Englund}%
\affiliation{Department of Electrical Engineering and Computer Science, Massachusetts Institute of Technology, Cambridge, MA 02139, USA}\affiliation{Research Laboratory of Electronics, Massachusetts Institute of Technology, Cambridge, MA 02139, USA}

\author{Matthew E. Trusheim}
\affiliation{Department of Electrical Engineering and Computer Science, Massachusetts Institute of Technology, Cambridge, MA 02139, USA}
\affiliation{Institute for Soldier Nanotechnologies, Massachusetts Institute of Technology, Cambridge, MA 02139, USA}
\affiliation{DEVCOM, Army Research Laboratory, Adelphi, MD, 20783, USA}

\date{\today}

\begin{abstract}
Nanomechanical structures have been investigated as a method of achieving long-lived quantum excitations at radio frequencies. Their high quality factors are especially intriguing as a medium for bosonic encoding of quantum information. However, to leading order, mechanical modes typically lack the nonlinearities necessary to achieve interaction between bosonic channels and thus are limited in their ability to scale to the many-qubit regime necessary for practical quantum computing. In this work, we propose and describe an approach for bosonic quantum information processing that uses strain-sensitive solid-state spins as nonlinear elements to produce the relevant nonclassical mechanical states. We outline the architecture required to achieve nearest-neighbor connectivity between mechanical cat-state qubits on-chip, as well as the control and readout architecture required for universal quantum computation. In addition, we show that this architecture can allow for a high spatial density of logical qubits by leveraging both the efficiency of bosonic error correction schemes and the small sizes of the constituent nanomechanical resonators and spin qubits. Finally, we identify the necessary performance metrics that will enable error-correction thresholds at high qubit densities, illuminating a path towards scalable quantum information processing.  




\end{abstract}

\maketitle

\section{Introduction}

Quantum information processing is predicated on using long-lived quantum modalities in an error-correctable computation scheme to achieve a quantum circuit. To date, large-scale efforts to achieve quantum error correction have focused on implementing surface codes\cite{google2023suppressing}, which have been shown to demonstrate sub-threshold error correction for increasing code distance, $d$\cite{google2025quantum}. However, surface codes are resource intensive, requiring $\sim d^2$ physical qubits per logical qubit, which causes the quantum information density to be low. Ultimately, such resource-intensive schemes will likely cause a bottleneck to scaling. An alternative approach is to perform bosonic encoding, where one leverages the infinite-dimensional Hilbert space of a harmonic oscillator to encode information. Typical approaches implement these bosonic modes in systems such as microwave electromagnetic cavities, in turn coupled to nonlinear superconducting circuits e.g. transmons to provide the requisite non-classical manipulation \cite{gao2019entanglement,putterman2025hardware}. These modalities have recently enabled quantum error correction schemes with logical information lifetimes greater than the physical lifetime of the mode\cite{ofek2016extending,sivak2023real}. However, using microwave circuitry is costly in size and the relevant decay timescales: the three-dimensional microwave cavities needed to achieve lifetimes greater than milliseconds are cm$^3$ in dimension, and the nonlinear ancilla transmons have typical dephasing times on the order of tens of microseconds. This leaves a narrow time window during which quantum operations can be performed on the bosonic cavity (since the transmon must be coherent during cavity operations) and limits prospects for scaling beyond the NISQ regime.

In contrast to the large dimensions required to house a microwave radiation mode, mechanical resonators in the GHz frequency range can be $\mu$m$^3$ in dimension. Indeed, the acoustic velocity in most materials is roughly four orders of magnitude smaller than the speed of light in a vacuum. Thus, a relatively large mechanical mode would occupy less space on-chip than a microwave transmission line resonator with lesser demonstrated lifetimes. Despite this, mechanical modes lack a native nonlinearity to first or second order in most materials, and thus preparation of a computationally useful nonclassical state is difficult. Piezoelectrics and superconducting transmons can be used to circumvent this issue, as has been demonstrated in the literature \cite{Bild2023}. However, transmons suffer in both size constraints and qubit lifetimes, limiting the efficacy of a space-efficient platform, and heterogeneous integration of superconductors and piezoelectrics into a low-loss mechanical system is an engineering challenge.

As an alternative system, we consider the prospect of using mechanical resonators coupled to strain-sensitive atomic defect spins in solid state as the ancillary qubits for quantum state preparation. Unlike transmons, the spin degrees of freedom of atomic defects are inherently long lived, with population lifetimes on the order of seconds, pure dephasing times surpassing 100 $\mu$s, and dynamically decoupled dephasing times approaching ms \cite{Karapatzakis2024}. Thus, a spin coupled to a mechanical resonator can allow for many more quantum operations on a mechanical mode within its decay time than a transmon would allow for an analogous system. This concept inverts previous proposals which use spins as the encoding qubit and mechanical modes as a mediating interaction, therefore confining logical quantum information to a much smaller device area \cite{kuzyk2018scaling, li2019honeycomblike,arrazola2024toward}. 

In this work, we develop a unit cell, consisting of a nanomechanical oscillator and ancilla defect spin, that capable of the full set of operations required for cat-qubit-based bosonic computation. We numerically demonstrate, via Lindblad and quantum Monte Carlo simulation of realistic device parameters, (i) dispersive preparation of even/odd cat states with fidelity exceeding 0.98; (ii) single-shot optical spin readout of cat-state parity with fidelity exceeding 0.95, used both to herald state preparation and to perform stabilizing parity checks; (iii) single-qubit Z and X logical gates implemented via dispersive phase accumulation and echoed conditional displacement, respectively; and (iv) spin-mediated two-qubit entangling operations, i.e. heralded Bell-pair generation and a CNOT gate between neighboring mechanical qubits with fidelities up to 0.98. Building an explicit error budget for the resulting parity-check cycle, we identify the spin and resonator coherence properties required to exceed a circuit-level error-correction threshold, and show that the long intrinsic coherence of diamond spin ancillae (rather than the mechanical quality factor) is a non-limiting resource in this budget, in contrast to transmon-based bosonic platforms where ancilla dephasing dominates. Because each logical qubit occupies only the micron-scale footprint of a single nanomechanical resonator, this architecture offers a path to information densities substantially exceeding those of microwave-cavity bosonic codes or superconducting surface codes (Table II), therefore confining logical quantum information to a much smaller device area \cite{kuzyk2018scaling, li2019honeycomblike,arrazola2024toward}. 

\begin{figure*}
    \centering
    \includegraphics[width=\linewidth]{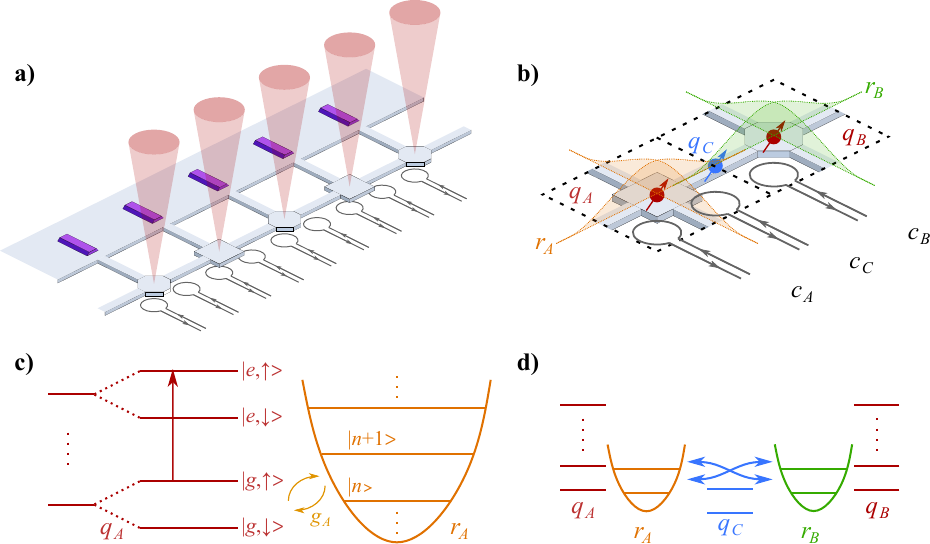}
    \caption{a) A depiction of the device. Piezoelectric transducers (purple) are driven by a microwave tone coupled via an interdigitated capacitor. Each spin can be optically addressed (depicted by red cones) for spin polarization and readout. b) A unit cell of the device which contains three qubits. A coupling qubit $q_c$ ties together each logical cell. A logical cell consists of a spin qubit and a nanomechanical resonator. Logical information is encoded in the bosonic states of the nanomechanical resonator, which can be driven by the piezo and addressed by the spin qubit. Superconducting wires flow current to control the spin qubit. c) Spin (red) and nanomechanical (yellow) energy diagrams. The spin has two optically addressed manifolds which can be split into two spin states. The spin is coupled with strength, $g$, to the nanomechanical resonator. d) Energy diagram of the full unit cell. By driving $q_c$ via the superconducting cables, one is able to perform two qubit operations between the two logical cells.}
    \label{fig:1_schematic}
\end{figure*}

\section{Demonstration and Performance of Bosonic Encoding}

We evaluate our platform against the DiVincenzo criteria for a viable quantum computing architecture \cite{DiVincenzo2000}, addressing, in order: (1) definition of physical qubit resources, (2) definition of a logical qubit from those resources, (3) qubit initialization, (4) readout of the logical qubit state, (5) a universal gate set, and (6) coherence times sufficient for fault-tolerant operation. We take each in turn below.

\subsection{Spin-Mechanical Bosonic Encoding Unit Cell}

The unit cell of our proposed architecture (Figure \ref{fig:1_schematic}a\&b) consists of (a) mechanical resonators $r_A$ and $r_B$ which will contain the logical cat states; (b) individually-coupled spins $q_A$ and $q_B$ that interact with solely their respective resonators at rates $g_{i=j}$, which will be used for single-qubit manipulation; (c) a coupler spin $q_C$ that couples to both resonators, for mediating two-qubit interactions.  

Each unit cell also includes several classical control elements. Two piezoelectric input ports coupled to $r_A$ and $r_B$, respectively, populate the phononic resonators via a small electromechanical interaction driven by a large, tunable driving field. Optical ports (indicated here as free-space coupling) allow measurement of the spin ancillae for effective non-destructive readout of the phononic resonators. A superconducting control loop for each ancillary spin, $c_{A,B,C}$, crucially allows tuning of the spin transitions. Quasi-static currents in the superconducting wires allow for tuning of the spin energy by up to $\sim300$ MHz via the Zeeman effect, limited by the critical current density of the superconductor (for Nb, $10^5-10^6$ A/cm$^2$)\cite{Huebener1975}, while resonant microwaves allow for spin state manipulation. These control elements: piezoelectric driving, optical spin readout, and superconducting control, give us the necessary functionality within a single unit cell.

The unit cell Hamiltonian, with energy levels labeled in Figure~\ref{fig:1_schematic}c-d, consists of the bare Hamiltonian $H_{cell,0}$ and interaction Hamiltonian $H_{cell,int}$
\begin{equation}
    \frac{H_{cell,0}}{\hbar} = \sum_{i=A,B}\omega_{R,i} a_i^\dag a_i + \sum_{i=A,B,C}\frac{\omega_{S,i}}{2}\sigma_{z,i};
\end{equation}

\begin{equation}
    \frac{H_{cell,int}}{\hbar} = \sum_{i=A,B} \sum_{j=A,B,C} (g_{x_{i,j}}\sigma_{x,j} + g_{z_{i,j}}\sigma_{z,j})(a_i + a_i^\dag).
\end{equation}

This interaction Hamiltonian can be rewritten in the rotating wave approximation (RWA) as the Jaynes Cummings Hamiltonian between each resonator and two-level system ancilla, 

\begin{equation}
    \frac{H_{cell,int}}{\hbar} = \sum_{i=A,B} \sum_{j=A,B,C} g_{i,j}(\sigma_{+,j}a_i + \sigma_{-,j}a_i^\dag).
\end{equation}

In the limit where the interaction between $r_A$ and $r_B$ is small due to low spatial mode overlap, the coupling terms $g_{A,B} \sim g_{B,A} \sim 0$. The crosstalk between $r_A$ and $r_B$ is further suppressed by the frequency mismatch between the resonant modes.

In the dispersive limit, when a given mode $i$ and ancilla $j$ are detuned by a frequency $\Delta_{i,j} >> g_{i,j}$, the interaction Hamiltonian can be expressed in the dispersive regime as

\begin{equation}
    H_{disp} = \frac{g_{i,j}^2}{\Delta_{i,j}}a^\dag_i a_i \sigma_{z,j} = \chi_{i,j}a^\dag_i a_i \sigma_{z,j}.
\end{equation}

For the parameter regime considered throughout this work (Table I), we take resonator frequencies $\omega_R/2\pi \approx 2$ GHz, spin transition frequencies $\omega_S/2\pi$ tunable over 300 MHz via the superconducting crossbar, spin–resonator coupling $g/2\pi \approx 1$ MHz, and typical dispersive detunings $\Delta/2\pi = 100$ MHz, giving dispersive shifts $\chi/2\pi = g^2/\Delta$ on the order of tens of kHz, which is well within the dispersive regime $\gamma \sim \Delta/100$ used for state preparation and readout in Sec. II.B–D, and consistent with the stronger coupling ($\chi/2\pi \approx 100-150$ kHz) required for sustained error correction in Sec. III.

Critically, the tunability of the spin ancillae allows us to move between the Jaynes-Cummings and dispersive regimes, providing full spin-mediated quantum control of the mechanical resonators as bosonic logical qubits.


\subsection{Definition, Initialization, and Measurement}

We now define our qubit states as bosonic encoded states in each resonator $r_{A/B}$, supported by quantum nonlinear operations enabled by $q_{A/B}$. Explicitly, we consider even and odd cat states, $|E> = \mathcal{N}_+\left(\ket{\alpha} + \ket{-\alpha}\right)$ and $|O> = \mathcal{N}_-\left(\ket{\alpha} - \ket{-\alpha}\right)$, with normalization $\mathcal{N}_{\pm} = 1/\left(\sqrt{2}\left(1 \pm e^{-2\abs{\alpha}^2}\right)\right)$ where, for mean photon number $\overline{n}$, $\overline{n}=\alpha^2$. We select cat states due to their resource-efficient error correction: in a coherent state logical basis $\ket{0}_L = \ket{\alpha}$, $\ket{1}_L = \ket{-\alpha}$, cat states are logical code words intrinsically error protected against photon loss due to the distinguishability of state parity \cite{cochrane1999macroscopically, mirrahimi2014dynamically}.

\begin{figure*}
    \centering
    \includegraphics[width=\linewidth]{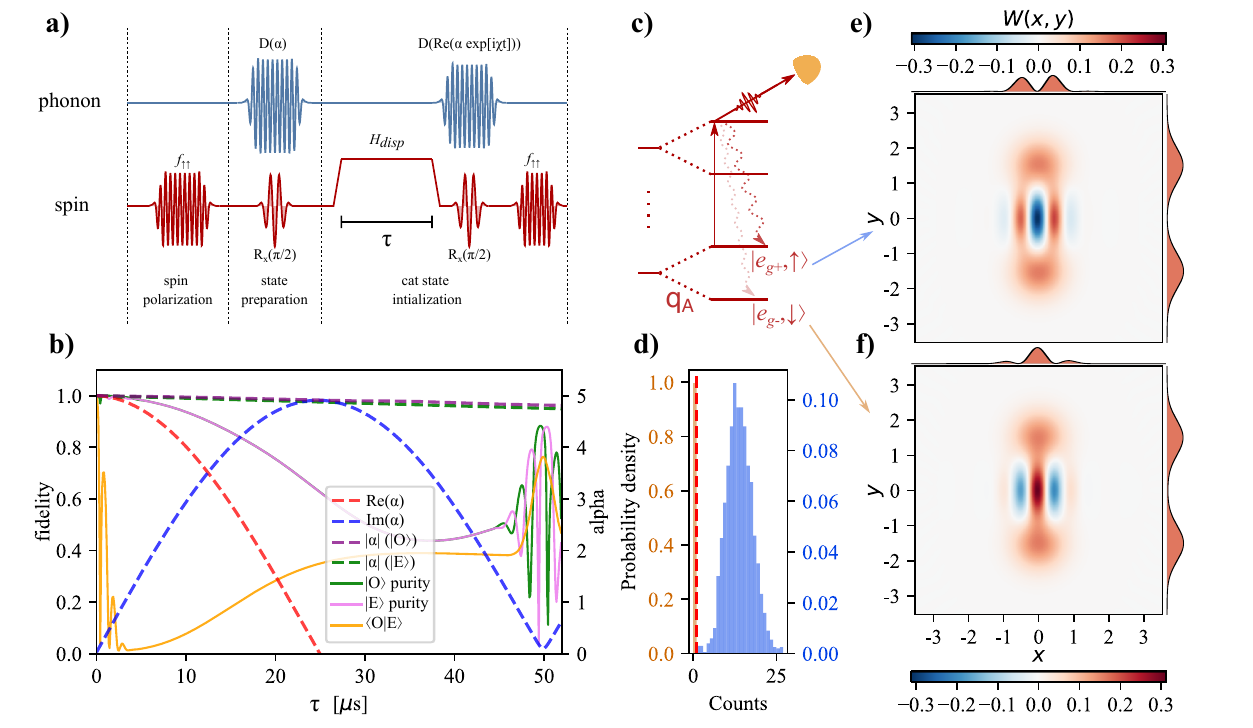}
    \caption{a) The cat state preparation scheme. b) Plot of the system evolution over time during dispersive interaction. The dotted red (blue) line represents the real(imaginary) displacement of the coherent state in quasi-probability space, where the real component is the amount by which the state must be shifted to achieve the initial states. The solid purple(green) lines represent the even(odd) cat state purities, while the solid purple(green) lines represent their magnitudes. The solid yellow line represents the orthogonality of the even/odd cat states. c) Schematic of single shot optical readout with d) readout histograms distinguishing the spin-up(orange) and spin-down(blue) states entangled with the odd and even cat states, respectively. The dotted red line at counts=1 represents the state distinguishing threshold. e) Odd cat state with negative Wigner contribution at the origin and therefore odd parity, heralded with the spin-up state. f) Even cat state with even parity, heralded with the spin-down state.}
    \label{fig:2-cat}
\end{figure*}

We begin by modeling the cat state initialization. We numerically model the evolution of the wavefunction of a single resonator--qubit pair under the influence of realistic decay channels using quantum Monte Carlo, with the parameters of Table~\ref{Tab:SimulationParams}. The procedure is illustrated in Figure~\ref{fig:2-cat}a and proceeds as follows.

First, we assume the piezoelectric transducer can initiate a coherent state of amplitude $\alpha$ (set to $5$ in Figure~\ref{fig:2-cat}) using an external drive at $\omega_{R,A/B}$~\cite{ArrangoizArriola2019}. Second, we assume the spin can be initialized to the superposition $\tfrac{1}{\sqrt{2}}\left(\ket{\downarrow} + \ket{\uparrow}\right)$, e.g.\ by optical pumping through external optics to polarize the spin state to $\ket{\downarrow}$~\cite{pingault2017coherent,trusheim2020transform}, followed by a $\pi/2$ rotation on the superconducting control line~\cite{Karapatzakis2024}.

Third, we tune the spin frequency to dispersively interact with the phononic resonator at rate $\chi = g^2/\Delta$, so that the system evolves roughly as
\begin{equation}
    \ket{\psi(t)} = \ket{\alpha e^{-i\chi t}}\ket{\uparrow}
    + e^{i\phi}\ket{\alpha e^{i\chi t}}\ket{\downarrow},
\end{equation}
where $\phi$ is the relative phase accumulated by the qubit ground state during unitary evolution. To perform this adiabatically starting from a well-decoupled spin (roughly $5$--$10\,\Delta$, see Table~\ref{Tab:SimulationParams}), the field must be tuned in under $\sim 25\ \mu$s. Figure~\ref{fig:2-cat}b plots the dispersive evolution of the resonator–spin pair under Eq. (5)–(7), including realistic decay, propagated numerically with the Lindblad master equation solver in QuTiP. Here, the $|\alpha|$ traces give the overall phonon loss of the system, while the state purities indicate progressive coupling of the resonator and spin. The $\text{Re}(\alpha)$ and $\text{Im}(\alpha)$ curves trace out the evolution of the resonator in phase space over the course of the dispersive evolution period.

Fourth, we displace the coherent state back to the origin and apply a $\pi/2$ rotation on the spin, generating
\begin{multline}
    \ket{\psi_f(t)} = \left(\ket{i \Im(\alpha e^{-i\chi t})}
    - e^{i\Phi}\ket{-i \Im(\alpha e^{-i\chi t})}\right)\ket{\uparrow} \\
    = i \left(\ket{i \Im(\alpha e^{-i\chi t})}
    + e^{i\Phi}\ket{-i \Im(\alpha e^{-i\chi t})}\right)\ket{\downarrow},
\end{multline}
where $\Phi = \alpha^2\sin(\chi t)$.

Finally, we read out the spin via resonant, spin-conserving optical excitation on the $\ket{g,\uparrow}\rightarrow\ket{e,\uparrow}$ transition (Figure~\ref{fig:2-cat}c). A spin projected into the addressed (``bright") state cycles the transition and emits many photons during the collection window, while a spin in the other (``dark") state is off-resonant and emits no photons during collection. Thresholding the resulting photon-count histogram (Figure~\ref{fig:2-cat}d) assigns the spin state in a single shot \cite{robledo2011high, rosenthal2024single}. Because the dispersive evolution has already entangled the spin with the parity of the mode, a direct single-shot optical readout heralds the positive or negative cat state,
\begin{equation}
    \ket{\mathcal{C}_{\pm}(\alpha)} =
    \ket{i \Im(\alpha e^{-i\chi t})} \pm e^{i\Phi}\ket{-i \Im(\alpha e^{-i\chi t})},
\end{equation}
with the spin-[$\uparrow$/$\downarrow$] outcome projecting the mode onto the even/odd cat, respectively.  We model this protocol as described using the Lindblad master-equation solver in QuTiP under realistic decay channels and find it is possible to initialize even and odd cat states of size $\alpha \sim 1.5$ (Figure~\ref{fig:2-cat}e,f) with fidelity $> 0.98$. We will also utilize this readout mechanism for mapping phonon-number parity for Wigner tomography (Sec. II.C), and reading out the joint-parity syndrome for two-qubit gates and the QEC cycle (Sec. II.D, III). Note that the coherent piezoelectric displacement to an initial $\alpha$ value, as well as the optical spin pumping, are assumed ideal and not modeled in this simulation. These idealizations are revisited quantitatively in the error budget of Sec. III, where finite $T_{1,s}$, $T_{2,s}$, and readout fidelity are considered closely.

\begin{table}
\begin{tabular}{| c | c | c | c| c| c| c|} 
 \hline
$Q_{\text{res}}$ & $T_{1,spin}$ (ms) & $g$ (MHz) & $f_{res}$ (GHz) & $\Delta$ (MHz) & $\alpha$ & $\eta$ \\ [0.5ex] 
 \hline
 10$^7$ & 1 & 1 & 2 & 100 & 5 & 1000\\
 \hline
\end{tabular}
\caption{Parameters used in the quantum Monte-Carlo Simulation. The parameter $\eta$ is the branching ratio used in QuTiP to simulate single shot optical readout of the spin ancilla \cite{sukachev2017silicon, gorlitz2022coherence}.}
\label{Tab:SimulationParams}
\end{table}

\subsection{Wigner Tomography via Single Shot Spin Readout}

Using the single-shot spin readout of Sec. II.B, we can additionally measure phonon-number parity following an arbitrary displacement, giving full Wigner tomography of the mechanical mode. The Wigner function fully characterizes the phonon mode, and its value at a phase-space point $\beta$ is proportional to the phonon-number parity of the state after it is displaced by $\beta$, $W(\beta) \propto \langle D^\dagger(\beta)\Pi D(\beta)\rangle$ with $\Pi = e^{i\pi a^\dagger a}$. Reconstructing $W$ therefore reduces to a single repeated operation: measuring phonon number parity after a controlled displacement. As we will explain, our spin can be used to measure the phonon number parity and can be probed via single-shot readout.

\begin{figure}
    \centering
    \includegraphics[width=\linewidth]{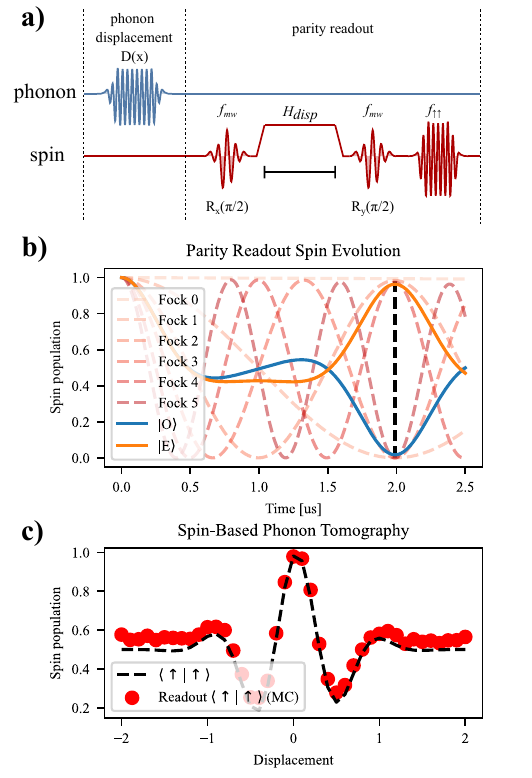}
    \caption{Spin-enabled wigner tomography of the phonon state. a) wigner tomography scheme. b) spin population evolution under dispersive interaction with the phonon for different phonon states, including the odd and even cat states prepared in Figure \ref{fig:1_schematic}. c) Wigner tomography with spin population averaged over 1000 Monte Carlo simulated tomography experiments with varying real displacement of the even cat state}
    \label{fig:3-cat_benchmarking}
\end{figure}

The parity-mapping sequence is shown in Figure~\ref{fig:3-cat_benchmarking}a. We first optically pump the spin into a definite ground state with a resonant laser at $f_{\uparrow\uparrow}$, driving the spin-conserving transition $\ket{e_{g-},\uparrow}$ and relying on the branching decay to polarize into $\ket{e_{g+},\downarrow}$. We then displace the phonon mode by $\beta$ and dispersively couple the spin to the mode for the parity time $t_p = \pi/2\chi$, which imprints a spin-dependent phase of $\pi$ per phonon and so maps the parity of the displaced state onto the spin: even parity leaves the spin in $\ket{\uparrow}$, odd parity in $\ket{\downarrow}$. A final $\pi/2$ pulse converts this phase into a population difference, which we read out optically.

\begin{figure}
    \includegraphics[width=\linewidth]{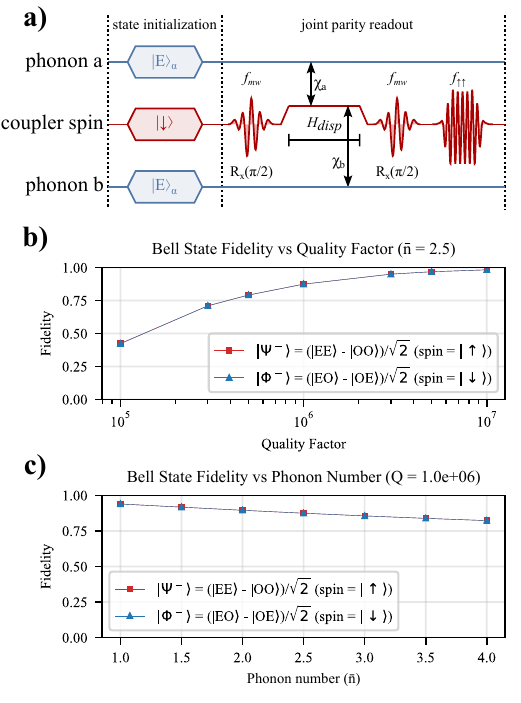}
    \caption{a) Parity readout-based entanglement scheme. b) Bell state fidelity for increasing mean phonon number, $\bar{n}$, in the resonator. c) Bell state fidelity for various $Q_{res}$ with $\bar{n}$ = 2.5.}
    \label{fig:4entanglement}
\end{figure}

We benchmark this single-shot optical parity readout following the protocol of Figure~\ref{fig:3-cat_benchmarking}a, defining the readout fidelity as
\begin{equation}
    F_{\text{readout}} = 1 - \tfrac{1}{2}\left(P(+|-) + P(-|+)\right),
\end{equation}
where $P(+|-)$ and $P(-|+)$ are the false-positive and false-negative probabilities. Using quantum Monte Carlo, we track the stochastic collapse of the spin into the optically ``bright'' and ``dark'' states that encode the parity outcome (Figure~\ref{fig:3-cat_benchmarking}b). Even parity leaves the spin in $\ket{g,\uparrow}$, which fluoresces brightly (high photon counts) under the $\ket{g,\uparrow}\!\to\!\ket{e,\uparrow}$ drive; odd
parity leaves it in the dark $\ket{g,\downarrow}$ state (low counts). By performing a single shot readout at time $t_p$ (dashed red line, Fig.~\ref{fig:3-cat_benchmarking}), we simulate a single-shot optical parity readout fidelity of $0.955$.

Sweeping $\beta$ and recording the spin signal reconstructs $W$; Figure~\ref{fig:3-cat_benchmarking}c shows the resulting fringes, whose oscillations reproduce the quantum interference pattern of the cat state. This same procedure underlies two further operations. For a state already entangled with the spin (as at the end of initialization) a direct spin readout heralds the cat parity without an additional mapping step. A parity check performed in situ enables stabilization: a detected parity jump signals a single phonon loss and triggers a feedforward displacement that returns the mode to the code space.

\subsection{Implementation of One- and Two-Qubit Gates}

We now describe the gate operations necessary to perform complete gate-based quantum computations on this platform. A full gate set for this quantum computing platform is represented by the gates:
\begin{itemize}
    \item Rotation about the x axis of the Bloch sphere, $R_x(\theta)$
    \item Rotation about the z axis of the Bloch sphere, $R_z (\theta)$
    \item a two-qubit CNOT gate.
\end{itemize}

Suppose that we select our logical states to be $\ket{0} = \ket{\alpha}$ and $\ket{1} = \ket{-\alpha}$. 
Then single-qubit $Z_L$ and $X_L$ logical gates are implemented via the coherent displacement and dispersive-phase shift protocols, respectively, described in the Supplement. We simulate their fidelities as a function of $\alpha$ and mechanical $Q$ factor in Figure S1, finding fidelities $\gtrsim 0.9$ for $Q \gtrsim 10^7$.

To perform a two-qubit operation, we require use of the ancillary spin $q_C$ between $r_A$ and $r_B$ to implement an entangling gate. This gate can be accomplished using a combination of a joint-parity readout and single-qubit gates. The joint-parity readout operation $P$ is as follows:

\begin{enumerate}
    \item Initialize $q_C$ in the superposition state $(\ket{\uparrow} + \ket{\downarrow})/\sqrt{2}$ with a $R_x(\pi/2)$ pulse.
    \item Tune $q_C$ so that it interacts dispersively with $r_A$ and $r_B$, with dispersive interaction rates $\chi_{A/B} = \frac{g_{A/B,C}^2}{\Delta_{A/B}}$.
    \item Allow the system to evolve under this dual dispersive coupling interaction.
    \item After time $\tau_{parity}$, apply a $R_x(\pi/2)$ pulse to $q_C$.
    \item Perform a spin readout on $q_C$.
\end{enumerate}
This dispersive interaction can be tuned by the ratio $\chi_{A}/\chi_{B}$, such that selected parity states result in ancillary $q_C$ spin-up and/or spin-down, allowing us to perform a control operation with $q_C$.

If the phonon number in $r_A$ and $r_B$ are jointly even (i.e. even/even, odd/odd), then $q_C$ will be in the $\ket{\uparrow}$ state; consequently, if  $r_A$ and $r_B$ are jointly odd (i.e. even/odd, odd/even), then $q_C$ will be in the $\ket{\downarrow}$ state. Measuring $q_C$, then, entangles $r_A$ and $r_B$ in the parity basis.

Figure~\ref{fig:4entanglement} shows numerical simulations of this heralded Bell pair generation protocol for realistic parameters. We find an expected decay in entanglement fidelity for increasing phonon number: importantly, for realistic $Q_{res}$ of $10^6-10^7$ and $\bar{n} = 2.5$, we find entanglement Uhlmann fidelities of 87\%-98\%. 

We note two observations: firstly, that the states $\ket{\Psi^-}$ and $\ket{\Phi^-}$ are identical up to a global phase after the application of an $X$ gate on $\ket{\Phi^-}$. Hence, the probabilistic Bell state generation can be made deterministic with a subsequent $X$ gate. Secondly, rewriting in the coherent state basis $\ket{0}_L = \ket{\alpha}$ and $\ket{1}_L = \ket{-\alpha}$ gives us $\ket{\Psi^-} = \left(\ket{\alpha,\alpha} - \ket{-\alpha,-\alpha}\right)/\sqrt{2}$, putting $r_A$ and $r_B$ in the coherent logical basis with intrinsic protection against single photon loss \cite{cochrane1999macroscopically}.

Furthermore, we can realize a CNOT gate by following a similar procedure to that of Figure~\ref{fig:4entanglement}. By entangling $q_C$ with the parity of $r_A$, we can impart a parity-conditional phase shift upon $r_B$, achieving a CNOT operation. The modified protocol follows the same steps 1-4 (interacting with $r_A$ $r_B$ simultaneously) and then performs:

\begin{enumerate}
    \item Tune $q_C$ so that it interacts dispersively with $r_B$, and perform a spin-conditional displacement upon $r_B$.
    \item Disentangle $q_C$ from $r_A$ by echoing the pulse sequence in steps 1-4.
\end{enumerate}

The net effect of this modified operation is a coherent $r_A$ parity-dependent CNOT operation on $r_B$, with similar fidelity scaling as that of Figure~\ref{fig:4entanglement}. This gives us all of the tools necessary to implement quantum computations on our proposed platform. 

\section{Coherence }
\begin{figure*}[t!]
    \centering
    \includegraphics[width=\linewidth]{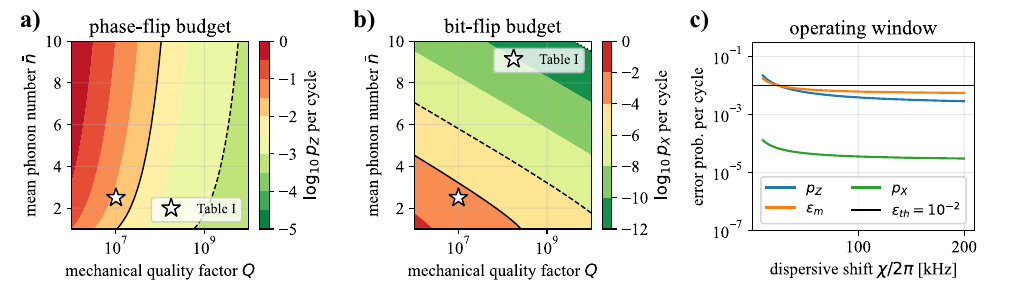}
    \caption{Error-correction parameter requirements. a) Per-cycle phase-flip probability $p_Z$ (Eq. 10) as a function of mechanical quality factor $Q$ and mean phonon number $\bar{n}$, with the Table I operating point marked (star) and the $\epsilon_{th} = 10^{-2}$ threshold contour (dashed). b) Per-cycle error probabilities — phase-flip $p_Z$ (blue), syndrome/measurement error $\epsilon_m$ (green), and bit-flip $p_X$ (orange) — as a function of dispersive shift $\chi/2\pi$, against the $\epsilon_{th} = 10^{-2}$ threshold (black). c) Per-cycle bit-flip probability $p_X$ (Eq. 12) as a function of $Q$ and $\bar{n}$, with the same Table I operating point and threshold contour as in (a).}
    \label{fig:threshold_analysis}
\end{figure*}
The four preceeding criteria establish the physical system, state preparation, readout, and a universal gate set. The final element is the relevant coherence time, which we take to be one that is sufficient for quantum error correction (QEC) and thus enables scalability. By constructing an error budget for the elementary QEC cycle, we can estimate device parameters to place each uncorrected error channel below a fault tolerance threshold.

\subsection{QEC Cycle}
In the coherent-state basis $\{\ket{\alpha}, \ket{-\alpha}\}$, in which the cat states are X eigenstates, the cat qubit exhibits the strongly biased noise underpinning repetition-cat architectures \cite{guillaud2019repetition}. A single phonon loss maps $\ket{\mathcal{C}_{\pm}}\rightarrow\ket{\mathcal{C}_{\mp}}$ and therefore acts as a logical phase flip with rate $\Gamma_{Z} \simeq \overline{n}\kappa(1+2n_{th}),$
with $n_{th}$ the thermal occupation of the mode and $\kappa = \omega_{res}/Q$. Logical bit flips require traversing the phase-space separation of the coherent state components and are thus exponentially suppressed with rate $\Gamma_X \sim \overline{n}\kappa e^{-2\overline{n}}$. At $f_{res} = 2$ GHz, the thermal occupation is negligible at dilution refrigerator base temperatures of $\sim 15$ mK; thus millikelvin operation is assumed throughout.

The elementary   cyle is a parity check (Sec. IIIB) followed by a feed-forward operation. A dispersive parity check lasts for time $t_{c} = t_{p} + t_{ro} + t_{rst}$, with parity time $t_{p} = \frac{\pi}{2\chi}$, single-shot optical spin readout time $t_{ro}$, and spin repolarization time $t_{rst}$.  Per cycle, the budget therefore incurs three possible errors. First, a phase-flip probability
\begin{equation}
    p_Z \simeq \overline{n}\kappa(1+2n_{th})t_c + \frac{t_p}{2T_{1,s}}
    \label{eq:p_Z}
\end{equation}
combines phonon jumps with ancilla spin relaxation of time constant $T_{1,s}$ during the parity check--an analogous relaxation to transmon-induced cavity dephasing \cite{sun2014tracking}. 

Next, the syndrome measurement error
\begin{equation}
    \epsilon_m \simeq (1 - F_{ro}) + \frac{t_p}{2 T_{2,s}}
    \label{eq:eps_m}
\end{equation}
consists of readout error with ancilla dephasing during the readout period. We note here that to minimize the syndrome measurement error, dynamical decoupling sequences will need to be interleaved with readout. In analogy to proposals for interleaved dynamical decoupling and parity readout in superconducting circuits \cite{Huembeli2017}, in the supplemental information we discuss the prospects of interleaving spin refocusing pulses and periods of spin detuning from the resonator before the final $\pi/2$ pulse of the readout sequence. This would decouple the spin ancilla from a surrounding bath, dramatically increasing $T_{2,s}$ from the pure dephasing time of the spin to a value closer to $T_{1,s}$, the cost of a small increase to the parity mapping operation time $t_p$.

Finally, the per-cycle bit flip probability
\begin{equation}
    p_X \simeq \overline{n}\kappa(1+2n_{th})t_c e^{-2\overline{n}},
    \label{eq:p_X}
\end{equation}
which the outer repetition code does not correct, remains exponentially suppressed \cite{mirrahimi2014dynamically}. We validated the dominant terms of Eqs.~\eqref{eq:p_Z} and ~\eqref{eq:eps_m} against Lindblad simulations of the full parity-check cycle using the model of Sec. IIIB, finding agreement at the few-percent level for a range of $10^6 < Q < 10^8$ and $10\;\mu\text{s} < T_{2,s}<10\text{ ms}$ (see Supplement).

For the concatenated repetition-cat code, phase flips are corrected provided the per-cycle data and syndrome errors lie below a circuit-level threshold of order $\epsilon_{th} \sim 10^{-2}$ \cite{guillaud2021error, chamberland2022building}. A convenient figure of merit is the number of parity checks per phase-flip time $t_c$,
\begin{equation}
    N_{\text{checks}} = \frac{1}{\Gamma_Z t_c},
\end{equation}
which is the acoustodynamic analog of the $\chi/\kappa$ cooperativity criteria of dispersive circuit QED \cite{blais2004cavity, blais2021circuit}. We would like $N_{checks} \gtrsim  1/\epsilon_{th} \approx 100$, which sets the parameter considerations below.

\subsection{Parameter requirements}
To reach a target of $N_{checks} \gtrsim100$, we consider, in order, (i) the required $g$ and $\Delta$ for sufficiently large $\chi$, (ii) the required resonator $Q_{min}$, and requisite ancilla lifetimes to meet this regime.

Figure~\ref{fig:threshold_analysis} shows the $p_Z$ (a) and $p_X$ (c) per cycle as a function of $Q$ and $\overline{n}$, with overall errors plotted as a function of dispersive coupling in (b). In Table 1, we consider a dispersive coupling of $10$ kHz to be well in the dispersive regime, $\gamma \sim \Delta/100$, which is sufficient for the state preparation, parity readout, and entanglement demonstrations of Sec. IIIA-C. However, sustained error correction benefits from $\chi \sim 100$ kHz, shown in Figure~\ref{fig:threshold_analysis}a and ~\ref{fig:threshold_analysis}b. The budget identifies the route across the threshold. Because $t_c$ is bounded below by the readout and reset times once $t_p \ll t_{ro}$, increasing $\chi$ beyond $\chi/2\pi \approx 100$ kHz saturates the required quality factor at
\begin{equation}
    Q_{min} \approx \frac{\overline{n}\omega_{res} (t_{ro} + t_{rst})}{\epsilon_{th}} \approx 2\times10^7.
\end{equation}

Reaching $\chi/2\pi=150$ kHz within the dispersive regime requires $g_{sm}/2\pi\approx3$ MHz at $\Delta/2\pi=60$ MHz--a threefold increase in coupling over Table I, which is still consistent with strain susceptibilities predicted for group-IV color centers \cite{meesala2018strain, Raniwala2025, joe2026purcell}. Alternatively, the resonator-ancilla detuning can be reduced to $\Delta \approx 20$ MHz, giving a marginal sub-threshold operation at $\chi/2\pi = 50$ kHz. Quality factors of order $10^7$ are well below the $Q\sim 10^{10}$ demonstrated in silicon photonic cyrstal cavities at millikelvin temperatures \cite{maccabe2020nano}; realizing these quality factors in diamond with embedded or integrated resonators remains the central fabrication challenge of this proposal.

The primary advantage revealed in this parameter estimation is that the spin ancilla is non-limiting in the system. Table 1 assumes a $T_{1,s} = T_{2,s} = 1$ ms, far below the coherence times demonstrated for SiV and GeV centers \cite{sukachev2017silicon, senkalla2024germanium}. In existing bosonic encoding platforms, the transmon ancilla ($T_{2} \sim 10\;\mu\text{s})$ limits the coherence budget, whereas the ancilla is enabling in our estimated thresholds. The default sequence simulated in our parity readout schemes (Fig.~\ref{fig:3-cat_benchmarking}a and Fig.~\ref{fig:4entanglement}a), where the ancilla or coupler spin dispersively interacts with the resonator(s) without any refocusing pulses, would not utilize the $T_{2,s}$ dynamical decoupling time, but rather the pure dephasing time $T_{2,s}^*$ of the ancilla; however, as already discussed, this can be remedied by interleaving periods of spin detuning from the resonator(s) alongside simultaneous spin refocusing pulses (see Supplement). The spin ancilla uniquely permits a "dynamically decoupled" parity readout scheme due to the long $T_{2,s}$ permitted under spin echo \cite{sukachev2017silicon, senkalla2024germanium}--an additional advantage built into this system.

The remaining syndrome channel is improved by increasing $F_{ro} \gtrsim 0.995$--greater than the estimated $F_{ro} = 0.955$--by similarly improving the resonator $Q_{res}$ to reduce coherence state decay during the parity-spin entanglement step in the readout sequence.

\begin{table*}[t]
\centering
\label{tab:footprint-comparison}
\begin{tabular}{@{}p{4cm}p{2cm}p{4.5cm}p{4.5cm}p{1.25cm}@{}}
\toprule
\textbf{System} & \textbf{Encoding} & \textbf{Footprint (area)} & \textbf{Status} & \textbf{Ref} \\
\toprule
This work (spin + nanomechanical cat) & bosonic & $\sim$10 $\mu$m$^2$ (single cat); $\sim$10$^2$ $\mu$m$^2$ incl. outer code & proposal & --- \\
\addlinespace
Planar cat + repetition code (AWS Ocelot) & bosonic & $\sim$12 mm$^2$ active / $\sim$1 cm$^2$ die ($d=5$) & demonstrated, below threshold & \cite{putterman2025hardware} \\
\addlinespace
3D cavity cat / GKP (Yale) & bosonic & $\sim$50 cm$^3$ (volume); cm$^2$ footprint & demonstrated, break-even+ & \cite{ofek2016extending,sivak2023real} \\
\addlinespace
HBAR + transmon & bosonic & $\sim$10$^{-3}$ mm$^3$ mode volume & cat-state demo & \cite{bild2023schrodinger} \\
\addlinespace
Transmon surface code (Google Willow, $d=7$) & surface & $\sim$0.1--1 cm$^2$ active / $\sim$4 cm$^2$ die & demonstrated, below threshold & \cite{google2025quantum} \\
\addlinespace
Neutral atoms ($d=7$ patch) & surface & $\sim$250 $\mu$m$^2$ trap region (excl. apparatus) & demonstrated, transversal logic & \cite{bluvstein2024logical} \\
\addlinespace
Majorana tetron (single, idealized) & topological & $\sim$15 $\mu$m$^2$ & roadmap projection & \cite{aasen2025roadmap} \\
\bottomrule
\end{tabular}
\caption{Approximate logical qubit area for different quantum platforms.}
\end{table*}

\section{Conclusion}
We have proposed and numerically modeled a platform for bosonic quantum computing that leverages nanomechanical resonators coupled to diamond vacancy center spins. We numerically showed that a single spin ancilla dispersively coupled to a phononic resonator is sufficient to prepare cat states with fidelities of $0.98$ for $\alpha = 1.5$. Single-shot optical readout of the spin ancilla enables high-fidelity parity measurement of the bosonic mode, with estimated readout fidelities exceeding $0.95$. By introducing a spin-bus protocol mediated by an ancillary coupling spin, we showed that two-qubit entangling gates between neighboring cat state qubits can be performed with fidelities up to $0.98$ at moderate phonon numbers and achievable resonator quality factors. 

Crucially, the $\mu m^2$ scale footprint of nanomechanical resonators offers a path toward information densities far exceeding those of microwave cavity or surface code architectures, while the long coherence times of diamond spins provide a wider operational window than transmon-based approaches. Table II situates this footprint against existing logical-qubit platforms: our proposed spin-resonator platform occupies roughly 10 $\mu\text{m}^2$, two to four orders of magnitude smaller than demonstrated planar-cat or 3D-cavity bosonic codes and roughly two orders of magnitude smaller than demonstrated transmon surface-code patches.

The two-component cat state encoding adopted here represents the minimal bosonic code our proposed platform supports with simulated fundamental building blocks--dispersive cat state preparation, single-shot parity readout, and Bell state generation--and mirrors the concatenated cat architectures currently leading in superconducting bosonic efforts \cite{reglade2024quantum, putterman2025hardware}. We note that refinements to the encoding scheme can be readily implemented on this architecture. For example, squeezing the cat basis states has been shown to extend bit-flip lifetimes by more than two orders of magnitude at fixed mean phonon number without increasing the phase-flip rate \cite{guillaud2023quantum, rousseau2025enhancing}, which would increase the time windows for spin-parity entanglement and readout and two-qubit operations. This squeezing can be generated by parametric modulation of the piezoelectric drive at $2\omega_{R_i}$ for resonator $r_i$ \cite{marti2024quantum}. Furthermore, the conditional $\sigma_z\left(a + a^\dag\right)$, which we abstracted from our system Hamiltonian, can be used to prepare and stabilize Gottesman–Kitaev–Preskill (GKP) grid states in our mechanical oscillators as a long-term option for logical qubits that have demonstrably reached break-even lifetimes \cite{sivak2023real}, autonomous stabilization \cite{lachance2024autonomous}, and universal gate sets \cite{matsos2025universal} on other hardware. The architecture is thus not tied to the coherent-state cat: it provides the nonlinearity, conditional displacements, and readout from which more robust hardware-efficient bosonic codes can be assembled.

\section{Acknowledgements}
HR acknowledges support from the MITRE Quantum Moonshot program. EGA acknowledges support from the Army Education Outreach Program Fellowship.

\section{Data Availability}
All of the data that support the findings of this study are reported in the main text and Supplementary Materials.  Source data are available from the corresponding authors on reasonable request.

\section{AI Disclosure}
Anthropic's Claude Sonnet 5 and OpenAI's GPT-5 were used for code, optimization and plotting. Furthermore, Claude and GPT-5 were used for polishing the manuscript and literature search.




\bibliography{references}%

\widetext
\newpage

\setcounter{figure}{0}
\renewcommand{\thefigure}{S\arabic{figure}}

\section*{Supplementary Material for Spatially Dense, Continuous-Variable Quantum Computing with Solid State Spin Nonlinearities}
\setcounter{section}{0}

\subsection{Description and Simulation of Single-Qubit Logical Gates}
In this section, we will describe single-qubit $X_L$ and $Z_L$ logical gates for the spin-mechanical oscillator unit cell. Let us assume that our logical states are chosen to be $\ket{0_L} = \ket{\alpha}$ and $\ket{1_L} = \ket{-\alpha}$. Then the parity operator $P = e^{i\pi \hat{a}^\dag\hat{a}}$ acts as a $X_L$ operation, and a photon loss acts as a $Z_L$ operation.

\begin{figure}
    \centering
    \includegraphics[width=\linewidth]{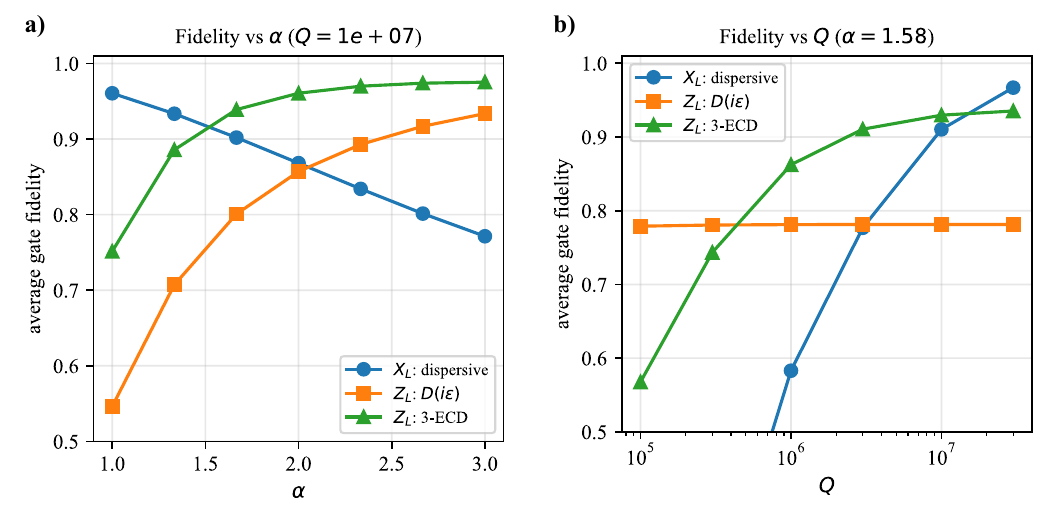}
    \caption{Single-qubit gate fidelities as a function of (a) resonator population $\alpha$ at fixed  $Q$ factor of $10^7$ and (b) $Q$ factor at fixed $\alpha$ of 1.58 from the main text.}
    \label{fig:single_qubit_gate_fidelities}
\end{figure}

We propose to achieve the $X_L$ gate by tuning the ancilla spin $q_i$ for a mechanical resonator $r_i$ to dispersively interact with rate $\chi = \frac{g^2}{\Delta}$ while the spin is in $\ket{g,\uparrow}$. Upon evolution time $t_X = \frac{\pi}{\chi}$, the unitary evolution under $H_{disp}$ becomes
\begin{equation}
    U(t_X) = e^{i \chi \hat{a}^\dag\hat{a}t} = e^{i \pi\hat{a}^\dag\hat{a}}=P.
\end{equation}
Note that this represents the action of the $q_i$ state on the phase of $r_i$, which is the analogous effect to the action of the $r_i$ parity on the phase of $q_i$ described in Section IIIB. The fidelity during this $X_L$ gate is limited primarily by photon loss in $r_i$; we plot the $X_L$ gate fidelity as a function of $\alpha$ and $Q$ in Figure~\ref{fig:single_qubit_gate_fidelities}. An alternative method of achieving a unitary "virtual $X_L$ gate" by re-framing the phase of subsequent $Z$ gates is also possible \cite{mckay2017efficient}.

To achieve an $Z_L$ gate, we propose two possible methods found in the literature. The naive method applies a small displacement via coherent drive to $r_i$ given by $D(i\epsilon)$, where $\epsilon = \frac{i\pi}{4\alpha}$ approximately implements $X_L$. The fidelity of this gate is limited by codespace leakage for finite $\alpha$. An alternative method involves applying an echoed conditional displacement (ECD) gate, which allows $r_i$ logical states to rotate with opposite angle in phase space under high $\alpha$ before returning them to the phase state origin \cite{eickbusch2022fast}. The procedure is roughly the following.
\begin{enumerate}
    \item Apply a displacement $D(r\alpha_0 e^{i\phi})$.
    \item Allow $r_i$ to evolve for a time $t_w$ under $H=\chi n\sigma_z$.
    \item Apply a displacement $D(-r_0\alpha_0 e^{i\phi})$
    \item Apply a spin flip $R_x(\pi)$.
    \item Apply a displacement $D(-r_1\alpha_0 e^{i\phi})$.
    \item Allow $r_i$ to evolve for $t_w$.
    \item Apply a displacement $D(r_2\alpha_0 e^{i\phi})$.
\end{enumerate}

Here, $\phi=\arg(\beta)+\pi/2$ and ratios $r=r_0=r_1=\cos(\chi t_w/2)$, $r_2=\cos(\chi t_w)$. The $t_w$ is calibrated so vacuum realizes the target conditional displacement $\beta$.

In Figure~\ref{fig:single_qubit_gate_fidelities}, we simulate the fidelities of $Z_L$ and $X_L$ gates, omitting the virtual gate which has unit fidelity. We plot the curves of three single-qubit gates: the $Z_L$, the $X_L$ implemented via small displacements, and the $X_L$ implemented with a sequence of spin rotations and ECD gates $R(\theta_3,\phi_3) ECD(\beta_2) R(\theta_2,\phi_2) ECD(\beta_1) R(\theta_1,\phi_1) ECD(\beta_0) R(\theta_1,\phi_1)$ with phase space angles $\theta_i, \phi_i$ and magnitudes $\beta_i$ optimized using a Nelder Mead simplex search algorithm.

Note that the ability to perform an arbitrary phase rotation on $r_i$ with the dispersive coupling interaction (i.e. $R_z(\theta)$ rotation) together with the $X_L$ gate allows for full Pauli operations on the $r_i$ logical qubit. Additionally, single-qubit gates on $r_i$ can be further optimized with protocols such as SNAP gates \cite{krastanov2015universal, heeres2015cavity} and GRAPE algorithm optimization of gates \cite{khaneja2005optimal}.

\section{Dynamically Decoupled Parity Readout}
\label{app:dd_chi}

The dispersive interaction Hamiltonian between $r_i$ and $q_i$, used for mapping the parity of a cat state encoded in $r_i$ onto the $q_i$ spin state, shares the same operator as dephasing noise on $q_i$:
\begin{equation}
    H_{disp,\delta} = \left(\chi\hat{a}^\dag\hat{a} + \delta\right)\sigma_z.
\end{equation}

Here, $\delta$ represents quasi-static noise experienced by $q_i$ due to a weakly interacting spin bath \cite{taminiau2012detection}. As such, during a normal dispersive interaction period, the spin $q_i$ dephases rapidly, preventing a high fidelity parity mapping. A naive dynamical decoupling protocol consisting of applying refocusing pulses in the form of $\sigma_x$ operators during the dispersive interaction period would cancel out phase accumulation due to the cavity. To see this explicitly, one can write $H_{disp,\delta}$ as a function of time with toggling function $f(t)$,
\begin{equation}
    H_{disp,\delta}(t) = f(t)\left(\chi\hat{a}^\dag\hat{a} + \delta\right)\sigma_z,
\end{equation}
where $f(t) = \pm 1$, flipping the sign of $\sigma_z$ after each refocusing pulse. Integrating, we find the total accumulated phase splits into a signal term and a noise term with identical structure,
\begin{equation}
\Phi_{\rm tot} = n \int_0^T f(t)\,\chi(t)\,dt \;+\; \delta \int_0^T f(t)\,dt,
\label{eq:phitot}
\end{equation}
where $n$ is the phonon number in $r_i$. For a balanced pulse sequence with $f(t) = 1$ and $f(t) = -1$ for equal time, both integrals vanish under the normal dispersive interaction period where $\chi(t)$ is a constant $\chi_0$ value. However, if $\chi(t)$ is driven synchronously with $f(t)$, then only the noise integral cancels and the desired phase accumulation may be nonzero.

As such, we propose a dynamically decoupled parity readout scheme. After a dispersive interaction time $t_h$ with $\chi = \chi_h$, apply a spin flip and detune $q_i$ further from $r_i$ such that $\chi = \chi_l \ll \chi_h$. Allow the system to evolve for time $t_l = 2 t_h$ and then apply a second spin flip and re-tune $q_i$ such that $\chi = \chi_h$. Allow the system to evolve for one more period $t_h$. Each evolution period results in the unitary operation $U_{h(l)} = e^{-i \chi_{h(l)}t_{h(l)}\hat{a}^\dag\hat{a} \sigma_z }$. The net operation on the $r_i$, $q_i$ subsystem becomes
\begin{equation}
    U_{parity} = U_h X U_l X U_h = e^{-2i\left(\chi_h - \chi_l\right)t_h\hat{a}^\dag\hat{a}\sigma_z}.
\end{equation}
The time $t_h$ can be selected such that $t_h = \frac{\pi}{4\left(\chi_h - \chi_l\right)}$ to achieve the desired parity mapping operation. In order to protect further against noise, one can interleave $2N$ refocusing pulses in the sequence, achieving the unitary
\begin{equation}
    U_{parity,N} = U_h \left(X U_l X U_h\right)^N = e^{-i\left((N+1)\chi_h - 2N\chi_l\right)t_h\hat{a}^\dag\hat{a}\sigma_z}.
\end{equation}

We can set $t_{l} = \frac{N + 1}{N}t_h$ so that quasi-static noise cancels in the toggling frame. Then the parity mapping operation is recovered when $t_h = \frac{\pi}{2(N+1)(\chi_h - \chi_l)}$ for a total evolution time $t_p = \frac{\pi}{\chi_h - \chi_l}$.

\begin{figure}
    \centering
    \includegraphics[width=\linewidth]{
    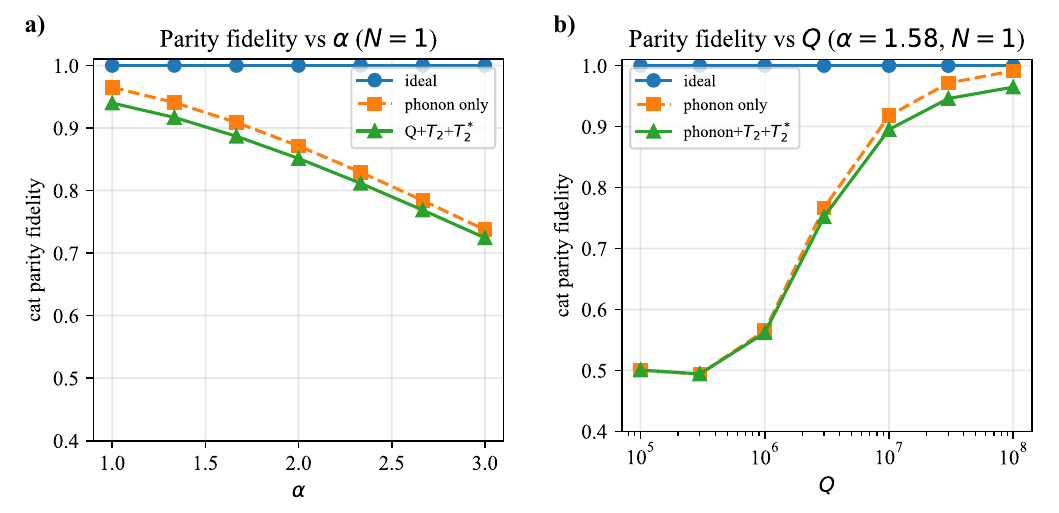}
    \caption{Parity mapping fidelities as a function of (a) resonator population $\alpha$ at fixed  $Q$ factor of $10^7$ and (b) $Q$ factor at fixed $\alpha$ of 1.58 from the main text. Setting $N=1$, or using two refocusing pulses, is sufficient in simulation to cancel quasi-static dephasing noise.}
    \label{fig:parity_mapping_sweep}
\end{figure}

Fig.~\ref{fig:parity_mapping_sweep} shows the scaling of parity mapping fidelity of $U_{parity,N}$ as a function of the $Q$ and $alpha$ of $r_i$, simulated using the Lindblad master equation solver in QuTiP with quasi-static noise injected in the system and $T_{2,s}$ set to 1 ms. We find that the spin dephasing due to quasi-static noise no longer limits the parity mapping fidelity, as desired. Practical implementations of the dynamically decoupled parity readout may interleave $XY$ pulses instead of only utilizing $X$ refocusing pulses to recover a $XY$ styled spin echo sequence \cite{gullion1990new}, as opposed to a Carr-Purcell-Meiboom-Gill styled spin echo \cite{carr1954effects,Meiboom1958}.

\end{document}